\documentclass[aps,pra,preprint,groupedaddress,showpacs]{revtex4}
\usepackage[dvips]{graphicx}

\begin{document}
\title{Loss of purity by wave packet scattering at low energies}
\author{Jia Wang, C. K. Law and M.-C. Chu}
\affiliation{Department of Physics, The Chinese University of Hong
Kong, Shatin, Hong Kong SAR, China}
\date{\today}
\begin{abstract}
We study the quantum entanglement produced by a head-on collision
between two gaussian wave packets in three-dimensional space. By
deriving the two-particle wave function modified by s-wave
scattering amplitudes, we obtain an approximate analytic
expression of the purity of an individual particle. The loss of
purity provides an indicator of the degree of entanglement. In the
case the wave packets are narrow in momentum space, we show that
the loss of purity is solely controlled by the ratio of the
scattering cross section to the transverse area of the wave
packets.
\end{abstract}
\pacs{03.67.Mn, 34.50.-s} \maketitle

In this Brief Report we describe the quantum entanglement
generated by wave packet scattering in three-dimensional free
space. Unlike one-dimensional problems studied previously by one
of us \cite{law}, scattering in 3D involves wave functions with
much richer state structures for entanglement. Recently, we have
demonstrated some interesting features for low energy
eigen-functions in trapped systems \cite{wang}. For unbounded
systems, Tai and Kurizki have analyzed the increase of entropy in
terms of the scattering matrix\cite{Kurizki}. Their approach is
based on a particular form of two-particle wave functions in which
the corresponding Schmidt decomposition can be expressed in pure
plane wave bases \cite{Kurizki}. For general two-particle wave
functions, however, particles may not be paired in plane wave
modes. Therefore a complete analysis of scattering effects on
entanglement production remains open for investigations. Here we
address the problem in the low-energy regime. Assuming the
interaction potential is isotropic and short-ranged, we can employ
the s-wave approximation to obtain the scattering wave functions.
Our task is to determine the loss of purity of an individual
particle, which serves as a measure of entanglement in our system
with pure two-particle states.

The system under investigation  consists of two interacting
particles of equal mass $m$ in free space. The Hamiltonian in
terms of center of mass and relative coordinates is given by: $H =
H_{cm} +  H_{rel}$ with
\begin{eqnarray}
&&  H_{cm} = \frac{{ P^2 }}{{2M}} \\ && H_{rel} = \frac{{ p^2
}}{{2\mu }} + V\left( { r} \right).
\end{eqnarray}
Here $M=2m$ is the total mass and $\mu=m/2$ is the reduced mass.
For convenience, we will use the units with $\hbar=\mu=1$. We
assume that the interaction potential $V\left( { r} \right)$ is
isotropic and has a short range $b$ such that $V\left( { r}
\right) \approx 0$ for $r>b$. Initially, the two particles are in
the form of (disentangled) gaussian wave-packets, each having a
width $\sigma_0$ in momentum space. Their initial positions and
average momenta are $\pm\mathbf{r}_0$ and $\mp\mathbf{k}_0$
respectively. The direction of ${\bf k}_0$ is chosen such that the
packets make a head-on collision at later time (Fig. 1).

\begin{figure}
\includegraphics[width=12.0cm]{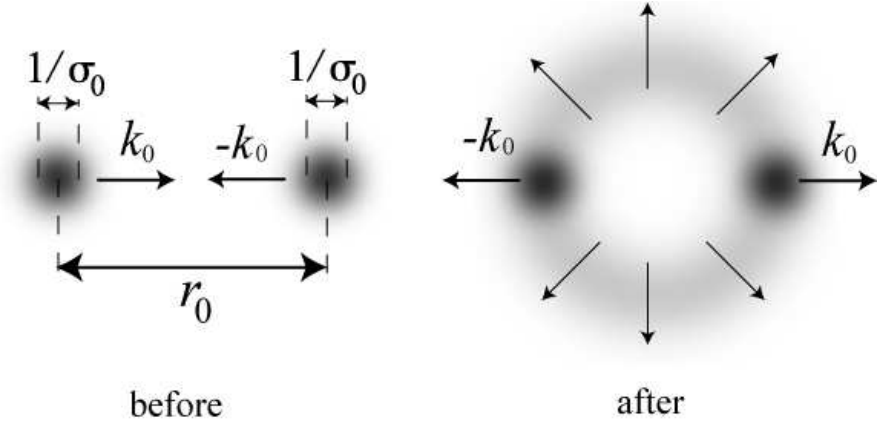}
\caption{An illustration of the system before and after a head-on
collision of wave packets. Under the s-wave approximation, the
scattered part of the single particle density (i.e., either
particle 1 or particle 2) is a spherical shell shown in the right
figure. The arrows indicate that the two particles go into
opposite directions.}
\end{figure}

The initial two-particle wave function in momentum space is given
by a product state: $\Phi \left( {\mathbf{k}_1 ,\mathbf{k}_2 ,0}
\right) = \phi _1 \left( {\mathbf{k}_1 } \right)\phi _2 \left(
{\mathbf{k}_2 } \right)$, where
\begin{eqnarray}
\phi _1 \left( {\mathbf{k}_1 } \right) =\Gamma\left( {\mathbf{k}_1
,\mathbf{k}_0 ;\sqrt 2 /\sigma _0 } \right) e^{ - i\left(
{\mathbf{k}_1  - \mathbf{k}_0 } \right) \cdot \frac{{\mathbf{r}_0
}}{2}}
\\
\phi _2 \left( {\mathbf{k}_2} \right) =\Gamma\left( {\mathbf{k}_2
,-\mathbf{k}_0 ;\sqrt 2 /\sigma _0 } \right) e^{ i\left(
{\mathbf{k}_2  + \mathbf{k}_0 } \right) \cdot \frac{{\mathbf{r}_0
}}{2}},
\end{eqnarray}
with $\Gamma\left( {\mathbf{a}, \mathbf{b} ;c } \right)$ being a
gaussian function parameterized by the inverse width $c$ and the
peak at $\mathbf{b}$,
\begin{equation}
\Gamma\left( {\mathbf{a}, \mathbf{b} ;c } \right) \equiv \left(
{\frac{{c ^2 }}{{2\pi }}} \right)^{3/4} \exp \left[ { - \frac{{c
^2 }}{4}\left( {\mathbf{a} - \mathbf{b} } \right)^2 } \right].
\end{equation}
The function (5) allows us to express the wave packets in a
compact form. After the scattering, the wave function in the long
time limit takes the form:
\begin{eqnarray}\label{Eq3}
\Phi \left( {\mathbf{k}_1 ,\mathbf{k}_2 ,t} \right) = \left( N
\right)^{ - 1/2} \left[ {\phi _1 \left( {\mathbf{k}_1} \right)\phi
_2 \left( {\mathbf{k}_2 } \right)  e^{-i(k_1^2+k_2^2) t/4} +
\varepsilon \phi_{scat} \left( {\mathbf{k}_1 ,\mathbf{k}_2,t }
\right)} \right].
\end{eqnarray}
Here the first term corresponds to a non-scattering part that
propagates freely, and the second term corresponds to the
scattering part. The constants $N$ and $\varepsilon$ are
normalization factors such that $\Phi$ and $\phi_{scat}$ are both
normalized to unity. In this paper we treat $|\varepsilon| \ll 1$
as a small number.

To analyze the quantum entanglement, it is customary to study the
entanglement entropy obtained from the Schmidt decomposition of
(6). Quite generally, the Schmidt modes are not simply the
momentum eigenfunctions, and the decomposition has to be performed
numerically. We note that this is in contrast to the special case
considered in Ref. \cite{Kurizki}, in which the Schmidt modes are
momentum eigenfunctions. To gain insight of the problem
analytically, we employ the purity function ${\cal P}$ as an
alternative measure of entanglement. Such a function is defined by
${\cal P} ={\rm Tr} (\rho_1 ^2)$, where $\rho_1 = {\rm Tr}_2
(\rho_{12})$ is the reduced density of the particle 1, and
$\rho_{12}$ corresponds to the two-particle density matrix
associated with the state (6). For pure two-particle states
considered in this paper, the smaller the value of ${\cal P}$, the
higher the entanglement. A disentangled (product) state
corresponds to ${\cal P}=1$. We remark that ${\cal P}$ shares
similar features as entropy, but it has the key advantage that it
is more accessible to theoretical analysis \cite{grobe,
Gemmer,chaotic}. In atomic physics,  ${\cal P}$ (or its inverse
${\cal P}^{-1}$) has also been employed to indicate the two-body
correlations in various dynamical processes \cite{grobe,liu}.

Specifically, ${\cal P}$ takes an integral form in our system:
\begin{eqnarray}
{\cal P} = \int {\int {\int {\int {\Phi \left( {\mathbf{k}_1
,\mathbf{k}_2 ,t} \right)\Phi \left( {\mathbf{k}_3 ,\mathbf{k}_4
,t} \right)} } } }  \Phi^{*}\left( {\mathbf{k}_1 ,\mathbf{k}_4 ,t}
\right)\Phi^{*} \left( {\mathbf{k}_3 ,\mathbf{k}_2 ,t} \right)d^3
\mathbf{k}_1 d^3 \mathbf{k}_2 d^3 \mathbf{k}_3 d^3 \mathbf{k}_4.
\end{eqnarray}
From Eq. (4) and Eq. (6), we obtain the expression of $P$ up to
the second order of $\varepsilon$:
\begin{equation}
{\cal P} \approx 1 - 2\left| \varepsilon  \right|^2 \left[ {1 +
I_1 - I_2 - I_3 } \right],
\end{equation}
where the integrals $I_1,I_2,I_3$ are defined by,
\begin{eqnarray}
&& I_1  = \left| {\int {\int {\phi_{scat} ^{*} \left(
{\mathbf{k}_1 ,\mathbf{k}_2 ,t} \right)\phi _1 \left(
{\mathbf{k}_1 } \right)\phi _2 \left( {\mathbf{k}_2 } \right)d^3
\mathbf{k}_1 d^3 \mathbf{k}_2 } } } \right|^2,\\ && I_2 = \int {
{\left| {\int {\phi_{scat} ^{*} \left( {\mathbf{k}_1 ,\mathbf{k}_2
,t} \right)\phi _1 \left( {\mathbf{k}_1 } \right)d^3 \mathbf{k}_1
} } \right|^2 } d^3 \mathbf{k}_2 },\\  && I_3  = \int { {\left|
{\int {\phi_{scat} ^{*} \left( {\mathbf{k}_1 ,\mathbf{k}_2 ,t}
\right)\phi _2 \left( {\mathbf{k}_2 } \right)d^3 \mathbf{k}_2 } }
\right|^2 } d^3 \mathbf{k}_1 }.
\end{eqnarray}
These integrals describe the interference between a non-scattered
wave and a scattered wave. It is interesting to note that there
are no first order terms in $\varepsilon$ in Eq. (8), as these
terms cancel each other once the $\varepsilon$ dependence in the
normalization constant $N$ is taken into account. We also remark
that as long as $\varepsilon$ is a small parameter, Eq. (8) is
valid for general two-particle states that are initially
separable, not just for gaussian wave packets.

To calculate $\varepsilon$ and $\phi_{scat}$, let us rewrite Eq.
(6) in terms of center of mass and relative coordinates: $ \Phi
\left( {\mathbf{k}_1 ,\mathbf{k}_2 ,t} \right) = \left( N
\right)^{ - 1/2} \phi _{cm} \left( {\mathbf{K},t} \right)\phi
_{rel} \left( {\mathbf{k},t} \right)$, where $\mathbf{K }=
\mathbf{k}_1 + \mathbf{k}_2$, $ \mathbf{k} = \left( {\mathbf{k}_1
- \mathbf{k}_2 } \right)/2 $, $\phi _{cm} \left( {\mathbf{K},t}
\right) = \Gamma \left( {\mathbf{K},0;1/\sigma _0 } \right)e^{ -
\frac{1}{8}K^2 t} $. Since we are interested in low-energy
scattering processes, we may keep only the s-waves of the
scattered part in $\phi _{rel}\left( {{\bf k},t} \right)$, i.e.,
\begin{equation}
\phi _{rel} \left( {{\bf k},t} \right) \approx \phi _{rel}^{NS}
\left( {{\bf k},t} \right) + \varepsilon \eta ^{(s)} \left( {{\bf
k},t} \right),
\end{equation}
where $\phi _{rel}^{NS} \left( {\mathbf{k},t} \right)  = \phi
_{rel} \left( {\mathbf{k},0} \right) e^{ - i\frac{1}{2}k^2 t}$ is
the freely propagating non-scattered part. In this way, we have
$\phi_{scat} ({\bf k}_1, {\bf k}_2,t)\approx \phi _{cm} \left(
{\mathbf{K},t} \right) \eta^{(s)}  ({\bf k},t)$.

Under the assumption that the two particles are well separated
(initially and finally) \cite{Eugen}, the scattering part
$\varepsilon \eta ^{(s)} \left( {{\bf k},t} \right)$ is given by,
\begin{equation}
\varepsilon \eta ^{(s)} \left( {{\bf k},t} \right) =
\frac{1}{{\left( {2\pi } \right)^{3} }} \int {\int {\phi _{rel}
\left( {\mathbf{k}',0} \right) f_0 \left( {k'}
\right)\frac{{e^{ik'r} }}{r} } } e^{  - i\mathbf{k} \cdot
\mathbf{r}- ik'^2 t/2} d^3 \mathbf{k}' d^3 \mathbf{r}
\end{equation}
where,
\begin{equation}
f_0 \left( {k } \right) = \frac{{e^{i2\theta \left( {k } \right)} -
1}}{{2ik }} \label{Eq5}
\end{equation}
is the s-wave scattering amplitude, and $\theta \left( {k }
\right)$ is the s-wave scattering phase shift. After some
calculations, we obtain,
\begin{equation}
\varepsilon \eta ^{(s)} \left( {\mathbf{k},t} \right) =
\frac{{\sigma _0^2 }}{{4k_0- i 2\sigma_0^2 r_0  }}\left[
{e^{2i\theta \left( k \right)} - 1} \right]\Gamma\left(
{\mathbf{k},k_0 \hat \mathbf{k};\frac{2}{{\sigma _0 }}}
\right)\frac{{e^{i\left( {k - k_0 } \right)r_0 } }}{k}e^{ -
i\frac{1}{2}k^2 t} . \label{Eq2}
\end{equation}
The constant $\varepsilon$ is determined from the norm of the
right side of Eq. (15). We may Taylor expand $\theta (k)$ at $k_0$
to the second order. The normalization condition is a Gaussian
integral that can be calculated explicitly. This gives
\begin{equation}
|\varepsilon|^2 = \frac{{\sigma _0^2 }}{{k_0^2\gamma^2}} \left[ {
1 - {\mathop{\rm Re}\nolimits} \left\{ {e^{2i\theta \left( {k_0 }
\right)} \sqrt {\frac{2}{{2 - i{\sigma _0^2}\theta ''\left( {k_0 }
\right)}}} \exp \left[ { - \frac{{\sigma _0^2 \theta '\left( {k_0
} \right)^2 }}{{ {2 - i\sigma _0^2\theta ''\left( {k_0 } \right)}
}}} \right]} \right\}} \right]\label{Eq4}
\end{equation}
where $\theta ' (k_0)$ and $\theta '' (k_0)$ are first and second
derivatives of $\theta  (k_0)$, and $\gamma^2 \equiv 1+
\left(\frac{\sigma_0^2 r_0}{2k_0}\right)^2$ in the denominator
corresponds to the spreading factor of the spatial width of the
packets (since $r_0/k_0$ is the time of collision). Therefore the
spreading of the wave packets would decrease the norm of the
scattered wave function $|\varepsilon|^2$ as expected. We also
note that the value of the bracket $\left[ {1 - {\mathop{\rm
Re}\nolimits} \left\{  \ldots \right\}} \right]$ is bounded
between 0 and 2, and therefore $|\varepsilon|^2$ is smaller than
$2 \sigma _0^2 /k_0^2 $.

With the results of $\varepsilon$ and $\phi^{(s)}$, we find that
$I_1 \approx \frac{{\sigma _0^2 }}{{2k_0^2 }},I_2 \approx
\frac{{2\sigma _0^2 }}{{3k_0^2 }},I_3 \approx \frac{{2\sigma _0^2
}}{{3k_0^2 }}$ are all of the order of $\sigma _0^2 / k_0^2 $.
Because of the prefactor $|\varepsilon|^2$ in Eq. (8), these
integrals's contribution to $P$ is about $ {{\sigma _0^4 /
}}{{k_0^4 }}$, which will be neglected.  Hence the purity of final
state is approximately
\begin{equation}
{\cal P} \approx 1 - 2\left| \varepsilon  \right|^2,
\end{equation}
where $\left| \varepsilon  \right|^2$ is given by Eq. (16).

Further simplification of this result can be made in the limit $
\sigma _0^2 \theta ''\left( {k_0 } \right) \ll 1 $, and $\sigma _0
\theta '\left( {k_0 } \right) \ll 1$, i.e., the wave-packets are
very narrow in the momentum space. In this limit, we have $\left|
\varepsilon  \right| \approx \sqrt 2 \left| {f_0 \left( {k_0 }
\right)} \right|\sigma _c$, where  $1/\sigma_c \equiv
\gamma/\sigma_0$ is the spatial width of the wave packets at the
collision time. Alternatively, we may employ the scattering cross
section $S_0 \left( {k_0 } \right)=4\pi |f_0 \left( {k_0 }
\right)|^2$, so that
 \begin{equation} 1- {\cal P} \approx 4 \sigma_c^2 |f_0 \left( {k_0 }
\right)|^2 =  \frac{{\sigma _c^2 S_0 \left( {k_0 } \right)}}{\pi}.
\end{equation}
Therefore the purity of the two-particle wave function after
scattering can now be explicitly expressed in terms of the s-wave
scattering cross section as well as the widths of wave packets.
However, we remark that such a simple relation is valid if
$\sigma_0$ is sufficiently small. The result can become more
complicated when $\sigma _0 \theta '\left( {k_0 } \right) $ or $
\sigma _0^2 \theta ''\left( {k_0 } \right) $ in Eq. (16) are not
negligible.

Equations (16-18) are the main results of this paper. We see that
the degree of entanglement (quantified by $1-{\cal P}$) is
determined by a simple dimensionless parameter $\sigma_c^2
S(k_0)$. Since $1/\sigma_c$ is the spatial width of an individual
wave packet at the collision time, $\sigma_c^2 S(k_0)$ is just the
ratio of scattering cross section to the characteristic cross area
of the wave-packet in position space. Hence, a stronger
entanglement can be generated for systems with a larger value of
the ratio. For example, this can be achieved by exploiting
resonance scattering in which $S(k_0)$ can be enhanced near the
resonance energies defined by the interaction
potential\cite{law,Kurizki}.

We point out that the degree of entanglement is typically small.
This is due to the fact that the two-particle wave function is
dominated by an un-scattered part, which is a product state.
However, if mainly the scattered part is observed (for example, by
detecting directions different from the incident one), then the
relevant wave functions can have a much higher degree of
entanglement. For the s-wave function given in Eq. (15), if
$\sigma_0$ is small such that the phase shift can be treated as a
constant $\theta(k_0)$, then the normalized (relative coordinate)
scattered wave function is a spherical shell of radius $k_0$ and
thickness $\sigma_0$ in momentum space. We find that the
corresponding purity function ${\cal P}$ has a leading term
proportional to $\sigma_0^2/k_0^2$ when $\sigma_0/k_0 \ll 1$ is a
small parameter. Therefore the narrower the width of the wave
packet, the stronger the entanglement in the scattered part of the
wave function.

To conclude, we present a simple and general formula that
approximates the loss of purity due to a head-on collision between
two gaussian wave packets in three dimensional space. As long as
the scattering is dominated by s-waves, our results provide a
quantitative measure of quantum entanglement generated. In
particular, our approach allows us to identify the key parameter
$\sigma_c^2 S(k_0)$, that explicitly connects the scattering cross
section and the width of wave packets to the degree of quantum
entanglement.

\begin{acknowledgments}
This work is supported in part by the Research Grants Council of
the Hong Kong Special Administrative Region, China (Project No.
400504).
\end{acknowledgments}

\end{document}